\documentclass[journal]{IEEEtran}

\usepackage{booktabs}
\usepackage{xcolor}
\usepackage{colortbl}
\usepackage{amssymb}
\usepackage{algorithmic}
\usepackage{amsmath} 
\usepackage{amsfonts}

\ifCLASSINFOpdf
\usepackage[pdftex]{graphicx}
\else
\usepackage[dvips]{graphicx}
\fi
\usepackage{epstopdf}
\usepackage{array}
\ifCLASSOPTIONcompsoc
\usepackage[caption=false,font=normalsize,labelfont=sf,textfont=sf]{subfig}
\else
\usepackage[caption=false,font=footnotesize]{subfig}
\fi
\usepackage{float} 
\usepackage{fixltx2e}
\usepackage{color}
\usepackage{xcolor}
\usepackage{textcomp}
\usepackage[normalem]{ulem}
\usepackage{cite}
\usepackage{multibib} 
\usepackage{nomencl}
\makenomenclature
\usepackage{etoolbox}
\renewcommand\nomgroup[1]{%
	\item[\bfseries
	\ifstrequal{#1}{P}{Physical Variables}{%
		\ifstrequal{#1}{M}{Mathematical Symbols}{%
			\ifstrequal{#1}{S}{Subscripts and Superscripts}{}}} ]}

\usepackage{hyperref}
\hypersetup{
	colorlinks=true,
	linkcolor=blue,
	filecolor=magenta,      
	urlcolor=cyan,}
\IEEEoverridecommandlockouts

\begin{document}
\bstctlcite{BSTcontrol}

\title{Robust Trajectory-Constrained Frequency Control for Microgrids Considering Model Linearization Error}

\author{Yichen~Zhang,~\IEEEmembership{Member,~IEEE,}
	~Chen~Chen,~\IEEEmembership{Senior Member,~IEEE,}
	~Tianqi~Hong,~\IEEEmembership{Member,~IEEE,}
	~Bai~Cui,~\IEEEmembership{Member,~IEEE,}
	~Bo~Chen,~\IEEEmembership{Member,~IEEE,}
	~Feng~Qiu,~\IEEEmembership{Senior Member,~IEEE}
	\thanks{
		This work was supported by U.S. Department of Energy Advanced Research Projects Agency - Energy through Network Optimized Distributed Energy Systems (NODES) Program.
		
		Y. Zhang, C. Chen, T. Hong, B. Chen and F. Qiu are with Argonne National Laboratory, Lemont, IL 60439 USA (email: thong@anl.gov).
		
		B. Cui is with National Renewable Energy Laboratory, Golden, CO 80401 USA.
		
}}

\maketitle

\begin{abstract}
Grid supportive (GS) modes integrated within converter-interfaced sources can improve the frequency response of renewable-rich microgrids.  However, the synthesis of GS modes to guarantee frequency trajectory constraints under a predefined disturbance set is challenging but essential.  To tackle this challenge, a numerical optimal control (NOC)-based control synthesis methodology is proposed. Without loss of generality, a wind-diesel fed microgrid is studied, where we aim to design GS functions in the wind turbine.  
In most control designs, linearized models are used. However, the linearization errors and proper compensation are not thoroughly investigated.
Here, linearized models are used, and the induced errors are quantitatively analyzed by reachability analysis and interval arithmetics, and represented in the form of interval uncertainties. Then, the NOC problem can be formulated into a robust mixed-integer linear program. The proposed control is verified on the modified 33-node microgrid with a full-order three-phase nonlinear model in Simulink. The simulation results show the effectiveness of the proposed control paradigm and the necessity of considering linearization-induced uncertainty.
\end{abstract}

%
\IEEEpeerreviewmaketitle

\mbox{}
\nomenclature[P,01]{All variables are in per unit unless specified.}{}
\nomenclature[P,02]{$\psi$}{Flux linkage}
\nomenclature[P,03]{$v$, $i$}{Instantaneous voltage, current}
\nomenclature[P,04]{$R$, $L_{l}$, $L_{m}$}{Resistance, leakage, mutual inductance}
\nomenclature[P,05]{$\overrightarrow{\Psi_{s}}$, $\Psi_{s}$}{Space vector of stator flux and its magnitude}
\nomenclature[P,06]{$\overrightarrow{V_{s}}$, $V_{s}$}{Space vector of stator voltage and its magnitude}
\nomenclature[P,07]{$H_{D}$, $H_{T}$}{Diesel, wind turbine generator inertia constant [s]}
\nomenclature[P,08]{$P_{m}$, $P_{v}$}{Diesel generator mechanical power, valve position}
\nomenclature[P,09]{$R$}{Governor droop setting}
\nomenclature[P,10]{$\tau_{d}$, $\tau_{g}$}{Diesel engine, governor time constant [s]}
\nomenclature[P,11]{$\omega_{c}$}{Cut-off frequency of low-pass filter [Hz]}
\nomenclature[P,11]{$\omega_{d}$, $\omega_{r}$}{Diesel, wind turbine angular speed}
\nomenclature[P,12]{$\omega_{f}^{*}$}{Filtered reference speed for wind turbine generator}
\nomenclature[P,13]{$\omega_{s}$}{Synchronous angular speed}
\nomenclature[P,14]{$\overline{\omega}$}{Speed base of wind turbine generator [rad/s]}
\nomenclature[P,15]{$\overline{f}$}{Speed base of diesel generator [Hz]}
\nomenclature[P,16]{$K^{T}$}{Torque controller gain}
\nomenclature[P,17]{$K^{Q}$}{Reactive power controller gain}
\nomenclature[P,18]{$K^{C}$}{Current controller gain}
\nomenclature[P,19]{$u_{\text{ie}}$}{Supplementary input for model reference control}

\nomenclature[M,01]{$A$, $B$, $E$}{State, control input, disturbance input matrices}
\nomenclature[M,02]{$C$, $D$, $F$}{Output, control feedforward, disturbance feedforward matrices}
\nomenclature[M,03]{$\Delta$}{Deviation from operating point}
	
\nomenclature[S,01]{$d$, $q$}{Direct, quadrature axis component}
\nomenclature[S,02]{$s$, $r$}{Stator, rotor}
\nomenclature[S,03]{$P$, $I$}{Proportional, integral}
\nomenclature[S,04]{$*$}{Reference and command}

\printnomenclature[0.6in]

\section{Introduction}\label{sec_introd}

\begin{figure*}[h]
	\centering
	\includegraphics[scale=0.3]{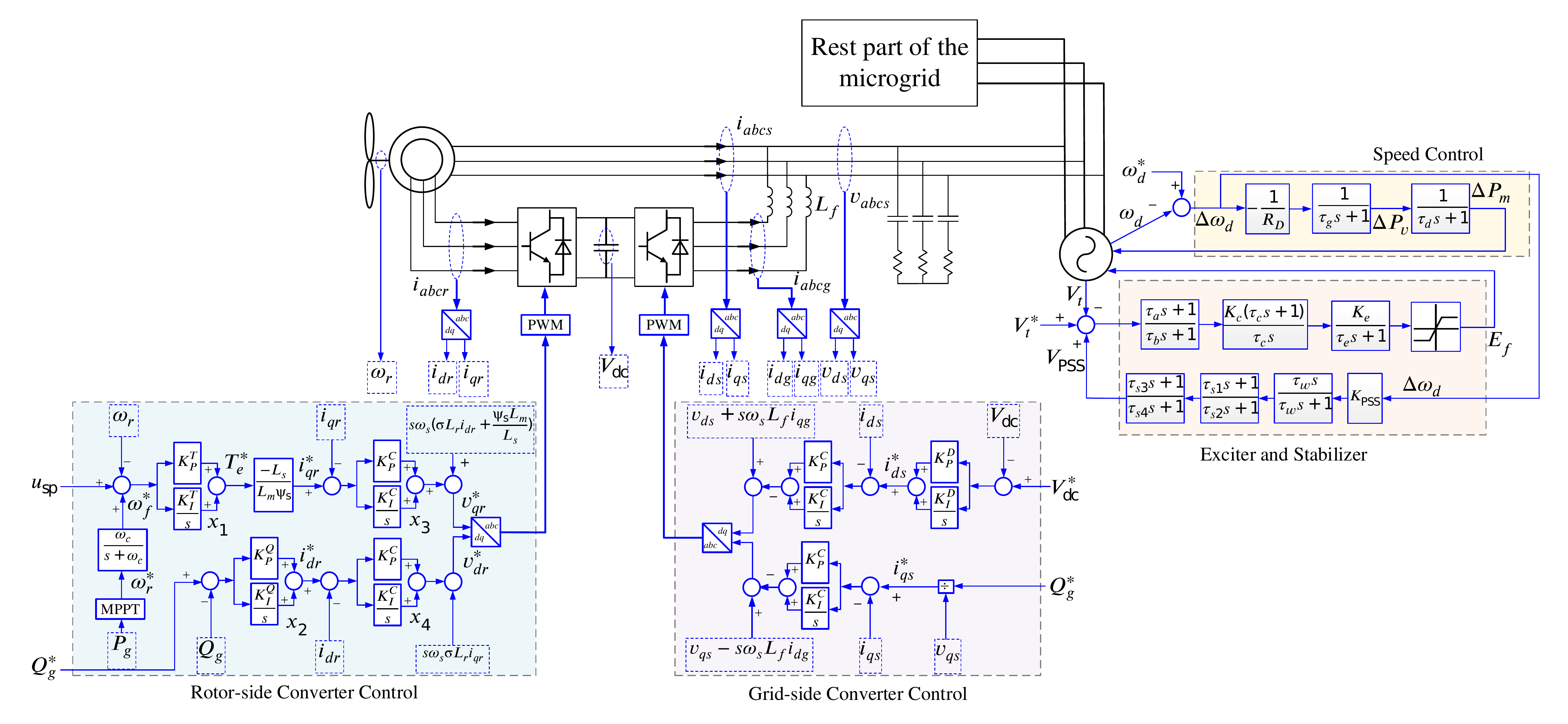}
	\vspace{-0.15in}
	\caption{A wind-diesel microgrid.}
	\label{fig_Model_Full}
\end{figure*}

Microgrids have become an ideal solution for powering remote locations, where wind and diesel units are among the most popular power sources \cite{allen2016sustainable}. Since the original converter control of wind turbine generators (WTGs) do not respond to frequency excursions, various grid supportive (GS) modes have been integrated to improve frequency response \cite{Bevrani2012,Serban2014,Han2015,Microgrid2015,Zhao2016,Wang2018a,Zhang2018a,Wang2019,Zheng2020,He2020}. However, little research has probed into the synthesis of GS modes to guarantee \emph{frequency trajectory constraints} under a predefined disturbance set.  Frequency transients that violate safe constraints may trigger unnecessary relay actions even though the system has adequate capacity to reach a viable steady state \cite{Soni2017}\cite{Pulendran2017}. But reaction strategies that can ensure a bounded frequency response given a disturbance is still not clear \cite{Uriarte2015}.

Despite the difficulty, several solutions have been proposed for this trajectory-constrained synthesis. In \cite{Uriarte2015}, given a disturbance, the available time remaining for resources to take actions to guarantee a bounded frequency response is estimated as a function of local inertia. An optimization-based commitment strategy for interruptible loads to ensure an adequate response is proposed in \cite{Bhana2016}. The frequency nadir information under different commitments of interruptible load needs to be obtained via simulation and sensitivity prediction. In \cite{Zhang2019}, the reachable set that ensures a safe trajectory is calculated and deployed as a supervisory control for GS mode activation. The gaps, however, still exist in the aforementioned works. First, all aforementioned approaches are based on pre-designed GS modes, and thus lack adequate flexibility for varying operating conditions. Second, although all studies employed linearized models, the linearization error and induced effects have not been rigorously analyzed and compensated.

We propose a novel robust control paradigm for regulating microgrid frequency. 
The objective is to preventively compute feasible supplementary control signals for WTGs so that the stored kinetic energy can be optimally used to support the frequency within permissible ranges. 
The control is strategically configured as a two-level paradigm. The upper level is a near real-time centralized control scheduling module, which solves a numerical optimal control (NOC) problem for GS control inputs. The lower level, which is configured in each WTG, stores the up-to-date control input in the normal condition, and executes the control input once a disturbance is detected. Such a preventive setting allows us to accommodate both trajectory constraints and control design flexibility.

To reduce the computation complex, linearized models are employed in the NOC problem. 
The present paper significantly extends our previous work \cite{Zhang2018b} by analyzing and compensating the linearization error. The reachability analysis and interval arithmetics are employed to calculate the linearization-induced uncertainty in the form of interval uncertainties. Then, the NOC is formulated as a robust mixed-integer linear program (MILP), the solution of which well compensates the linearization error.

The contributions of this paper are briefly concluded:
\begin{enumerate} 
	\item We address the challenging trajectory constraints in the frequency control and propose a novel two-level control paradigm.
	\item We analyze the linearization error of WTG models and obtain the induced bounds using Zonotope-based reachability analysis and interval arithmetics.
	\item We formulate the near real-time NOC problem as a robust MILP, analyze the worst-case realizations of both linearization-induced uncertainty and the disturbance set.
	\item We implement the proposed control paradigm with full-order three-phase nonlinear models in Simulink and verify its effectiveness.
\end{enumerate}

The remainder of the paper is organized as follows. 
Section \ref{sec_model} discusses the frequency response in microgrids and derives the linearization-induced uncertainty. 
Section \ref{sec_method} details the control paradigm. 
Section \ref{sec_sim} presents the case studies, followed by conclusion in \ref{sec_conclusion}.

\subsubsection{Notations}
Let $\dot{x}\in f(x)$ denotes differential inclusions. Let $\oplus$ denotes Minkowski addition of two sets.

\section{Microgrid Models}\label{sec_model}
The wind-diesel microgrid model is shown in Fig. \ref{fig_Model_Full}. The main objective of this section is to derive an augmented frequency response (AFR) model, which describes the microgrid frequency dynamics subjected to both disturbances and supports. Such models have been shown to be crucial for frequency studies \cite{Shi2018}. Compared with the models in \cite{Zhang2018a}, we use a full-order linearized WTG model instead and express the linearization error as an unknown-but-bounded set.

\subsection{Diesel Generator and Its Analytical Model}\label{sec_model_sub_DG}
A diesel generator (DG) is a combustion engine driven synchronous generator. A complete model consists of a two-axis synchronous machine, combustion engine, governor, and exciter shown in Fig. \ref{fig_Model_Full}. The governor, engine, and swing dynamics shown in (\ref{eq_SFR}) are extracted to describe the frequency characteristics of the diesel generator, which has proved to be precise in many power system applications \cite{Shi2018}
\begin{align}
\label{eq_SFR}
\begin{aligned}
2H_{D}\Delta\dot{\omega}_{d}&=\overline{f}(\Delta P_{m}-\Delta P_{e})\\
\tau_{d}\Delta\dot{P}_{m}&=-\Delta P_{m}+\Delta P_{v}\\
\tau_{g}\Delta\dot{P}_{v}&= -\Delta P_{v}  - \Delta\omega_{d}/(\overline{f}R_{D})
\end{aligned}
\end{align}
It is worth mentioning that the model in (\ref{eq_SFR}) can also present the frequency response of an aggregated group of DGs. Ref. \cite{Shi2018} and \cite{Zhang2018a} have shown such aggregation is accurate, especially in microgrids due to the closer electric distance. 

\subsection{Double Fed Induction Generator (DFIG)-Based WTG and Its Analytical Model}\label{sec_model_sub_WTG}
The zero-axis DFIG-based WTG can be described using the following differential-algebraic equations
\begin{align}
&\dot{\omega}_{r}=1/(2H_{T})(T_{m}-T_{e})\label{eq_IM_ode5}\\
&\dot{\omega}_{f}^{*}=\omega_{c}(\omega^{*}_{r}-\omega_{f}^{*})\label{eq_RSC_control_ode1}\\
&\dot{x}_{1}=K_{I}^{T}(\omega_{f}^{*}-\omega_{r}+u_{\text{sp}})\label{eq_RSC_control_ode2}\\
&\dot{x}_{2}=K_{I}^{Q}(Q^{*}_{g}-Q_{g})\label{eq_RSC_control_ode3}\\
&\dot{x}_{3}=K_{I}^{C}(i_{qr}^{*}-i_{qr})\label{eq_RSC_control_ode4}\\
&\dot{x}_{4}=K_{I}^{C}(i_{dr}^{*}-i_{dr})\label{eq_RSC_control_ode5}\\
&0= \overline{\omega}(v_{qs} - R_{s}i_{qs} - \omega_{s}\psi_{ds})\label{eq_IM_ode1}\\
&0= \overline{\omega}(v_{ds} - R_{s}i_{ds} + \omega_{s}\psi_{qs})\label{eq_IM_ode2}\\
&0= \overline{\omega}[v_{qr} - R_{r}i_{qr} - (\omega_{s}-\omega_{r})\psi_{dr}]\label{eq_IM_ode3}\\
&0= \overline{\omega}[v_{dr} - R_{r}i_{dr} + (\omega_{s}-\omega_{r})\psi_{qr}]\label{eq_IM_ode4}\\
&0=-\psi_{qs}+L_{s}i_{qs} + L_{m}i_{qr}\label{eq_IM_alg1}\\
&0=-\psi_{ds}+L_{s}i_{ds} + L_{m}i_{dr}\label{eq_IM_alg2}\\
&0=-\psi_{qr}+L_{r}i_{qr} + L_{m}i_{qs}\label{eq_IM_alg3}\\
&0=-\psi_{dr}+L_{r}i_{dr} + L_{m}i_{ds}\label{eq_IM_alg4}\\
&0=P_{g}+(v_{qs}i_{qs}+v_{qs}i_{qs}) + (v_{qr}i_{qr}+v_{qr}i_{qr})\label{eq_IM_alg5}\\
&0=Q_{g}+(v_{qs}i_{ds}-v_{ds}i_{qs}) + (v_{qr}i_{dr}-v_{dr}i_{qr})\label{eq_IM_alg6}\\
&\begin{aligned}\label{eq_RSC_control_alg1}
&0=-v_{qr}+x_{3}+K_{P}^{C}(i_{qr}^{*}-i_{qr})\\
&\qquad\qquad\qquad+(\omega_{s}-\omega_{r})(\sigma L_{r}i_{dr}+\frac{\Psi_{s}L_{m}}{L_{s}})
\end{aligned}\\
&\begin{aligned}\label{eq_RSC_control_alg2}
&0=-v_{dr}+x_{4}+K_{P}^{C}(i_{dr}^{*}-i_{dr})\\
&\qquad\qquad\qquad-(\omega_{s}-\omega_{r})\sigma L_{r}i_{qr}
\end{aligned}\\
& 0=-i_{qr}^{*} + \frac{-L_{s}}{L_{m}\Psi_{s}}[x_{1}+K_{P}^{T}(\omega^{*}_{f}-\omega_{r}+u_{\text{sp}})]\\
& 0=- i_{dr}^{*} + x_{2}+K_{P}^{Q}(Q^{*}_{g}-Q_{g})
\end{align}
The time scale of converter regulation compared to the frequency response is small enough to be neglected such that $v_{qr}=v_{qr}^{*}$ and $v_{dr}=v_{dr}^{*}$. Then, the loop is closed by the algebraic relations in (\ref{eq_RSC_control_alg1})-(\ref{eq_RSC_control_alg2}). The variables $u_{\text{sp}}$ and $Q^{*}_{g}$ are control inputs while $v_{ds}$ and $v_{qs}$ are terminal conditions. Due to the page limit, detailed description is omitted here and can be found in \cite{Zhang2018a}.

\subsection{Linearization Error and Induced Uncertainty}\label{sec_model_sub_WTG_lin}
As we can see, the frequency response of DGs in (\ref{eq_SFR}) is originally linear, while the WTG model from (\ref{eq_IM_ode5}) to (\ref{eq_RSC_control_alg2}) is nonlinear. To admit a MILP, linearization is taken, while the induced error is evaluated and expressed as a bounded uncertainty set. Let the overall WTG model in (\ref{eq_IM_ode5}) -- (\ref{eq_RSC_control_alg2}) expressed compactly as follows
\begin{equation}\label{eq_DAE}
\begin{aligned}
&\dot{x}=f(x_w,u_{\text{sp}},y_w),\quad 0=g(x_w,u_{\text{sp}},y_w)
\end{aligned}
\end{equation} 
with the output equations
\begin{equation}\label{eq_OUT}
\begin{aligned}
&z=h(x_w,u_{\text{sp}},y_w)
\end{aligned}
\end{equation} 
where $x_{w}=[\omega_{r},\omega^{*}_{f},x_{1},x_{2},x_{3},x_{4}]^{T}$, $z=[P_{g}]$, $y_{w}=[\psi_{qs},\psi_{ds},\psi_{qr},\psi_{dr},i_{qs},i_{ds},i_{qr},i_{dr},V_{qr},V_{dr},P_{g},Q_{g}]^{T}$.
Let $s=[x_w,u_{\text{sp}},y_w]$ and $s^{\text{eq}}=[x^{\text{eq}}_w,0,y^{\text{eq}}_w]$ denote the equilibrium point.
Using Taylor expansion at the equilibrium point $x^{\text{eq}}_{w}$ and $y^{\text{eq}}_{w}$ for the nonlinear DAEs yields
\begin{align}
\begin{aligned}
&\dot{x}_{w}\in f(s^{\text{eq}}) + A_{\text{sys}}\underbrace{(x_{w}-x^{\text{eq}}_{w})}_{\Delta x_{w}} + B_{\text{sys}}u_{\text{sp}} \\ &\quad\quad+C_{\text{sys}}\underbrace{(y_{w}-y^{\text{eq}}_{w})}_{\Delta y_w} \oplus I(\xi)\\
\end{aligned}\\
\begin{aligned}
&0 \in  \underbrace{g(s^{\text{eq}})}_{=0} + D_{\text{sys}}\underbrace{(x_{w}-x^{\text{eq}}_{w})}_{\Delta x_{w}} + E_{\text{sys}}u_{\text{sp}} \\ &\quad\quad+F_{\text{sys}}\underbrace{(y_{w}-y^{\text{eq}}_{w})}_{\Delta y_w} \oplus J(\xi)\\
\end{aligned}\\
\begin{aligned}
&z\in h(s^{\text{eq}}) + L_{\text{sys}}\underbrace{(x_{w}-x^{\text{eq}}_{w})}_{\Delta x_{w}} + M_{\text{sys}}u_{\text{sp}} \\ 
&\quad\quad + N_{\text{sys}}\underbrace{(y_{w}-y^{\text{eq}}_{w})}_{\Delta y_w} \oplus K(\xi)\\
\end{aligned}
\end{align}
where $I(\xi)$, $J(\xi)$ and $K(\xi)$ are the Lagrangian remainders.  The $i$th row of $I(\xi)$, $J(\xi)$ and $K(\xi)$, denoted as $I_i(\xi)$, $J_i(\xi)$ and $K_i(\xi)$, respectively, reads
\begin{align}
\begin{aligned}
I_i(\xi)=\frac{1}{2}(s - s^{\text{eq}})^{T}H^{d}_i(\xi)(s - s^{\text{eq}})\\
J_i(\xi)=\frac{1}{2}(s - s^{\text{eq}})^{T}H^{a}_i(\xi)(s - s^{\text{eq}})\\
K_i(\xi)=\frac{1}{2}(s - s^{\text{eq}})^{T}H^{o}_i(\xi)(s - s^{\text{eq}})\\
\end{aligned}
\end{align}
where $H^{d}_i$, $H^{a}_i$ and $H^{o}_i$ are the Hessian matrices for the $i$th row of the differential equation $f_{i}$, algebraic equation $g_{i}$, and output equation $h_{i}$.
Lagrangian remainders enclose all higher-order terms if $\xi$ can take any value of the linear combination of $s$ and $s^{\text{eq}}$
\begin{align}\label{eq_xi_set}
\xi\in\{\alpha s + (1-\alpha)s^{\text{eq}}|\alpha\in[0,1]\}
\end{align}

For index-1 DAEs, the algebraic variables can be expressed in terms of differential and control variables as follows
\begin{align}
\begin{aligned}
(y_{w}-y^{\text{eq}}_{w}) \in & - F_{\text{sys}}^{-1}D_{\text{sys}}(x_{w}-x^{\text{eq}}_{w})\\
&- F_{\text{sys}}^{-1}E_{\text{sys}}u_{\text{sp}}
\oplus [- F_{\text{sys}}^{-1}J(\xi)]\\
\end{aligned}
\end{align}
Then, the differential equations read
\begin{align}
\begin{aligned}
&\dot{x}_{w}\in f(s^{\text{eq}})+ [A_{\text{sys}} - C_{\text{sys}}F_{\text{sys}}^{-1}D_{\text{sys}}] (x_{w}-x^{\text{eq}}_{w}) \\
&\quad\quad+ [B_{\text{sys}} - C_{\text{sys}}F_{\text{sys}}^{-1}E_{\text{sys}}] u_{\text{sp}}\\
&\quad\quad \oplus I(\xi) \oplus [-C_{\text{sys}}F_{\text{sys}}^{-1}J(\xi)]
\end{aligned}
\end{align}
And, the output equations can be re-written as follows
\begin{align}
\begin{aligned}
&z\in h(s^{\text{eq}}) + [L_{\text{sys}} - N_{\text{sys}}F_{\text{sys}}^{-1}D_{\text{sys}}] (x_{w}-x^{\text{eq}}_{w}) \\
&\quad\quad + [M_{\text{sys}} - N_{\text{sys}}F_{\text{sys}}^{-1}E_{\text{sys}}] u_{\text{sp}}\\
&\quad\quad \oplus K(\xi) \oplus [- C_{\text{sys}}F_{\text{sys}}^{-1}J(\xi)]
\end{aligned}
\end{align}
Then, the equations expressed in terms of the deviation variables are expressed as follows
\begin{align}
\label{eq_WTG_linear_1}
\begin{aligned}
&\Delta\dot{x}_{w}\in \underbrace{[A_{\text{sys}} - C_{\text{sys}}F_{\text{sys}}^{-1}D_{\text{sys}}]}_{A_w} \Delta x_{w}\\
&\quad\quad+ \underbrace{[B_{\text{sys}} - C_{\text{sys}}F_{\text{sys}}^{-1}E_{\text{sys}}]}_{B_w} u_{\text{sp}}\\
&\quad\quad \oplus I(\xi)\oplus[-C_{\text{sys}}F_{\text{sys}}^{-1}J(\xi)]
\end{aligned}
\end{align}
\begin{align}
\begin{aligned}
&\Delta z\in\underbrace{[L_{\text{sys}} - N_{\text{sys}}F_{\text{sys}}^{-1}D_{\text{sys}}]}_{C_w} \Delta x_{w} \\
&\quad\quad + \underbrace{[M_{\text{sys}} - N_{\text{sys}}F_{\text{sys}}^{-1}E_{\text{sys}}]}_{D_w} u_{\text{sp}}\\
&\quad\quad \oplus K(\xi) \oplus [- C_{\text{sys}}F_{\text{sys}}^{-1}J(\xi)]
\end{aligned}
\end{align}
which can be simplified as follows
\begin{align}
\label{eq_WTG_linear_2}
\begin{aligned}
&\Delta\dot{x}_{w}\in A_w \Delta x_{w} + B_w u_{\text{sp}} \oplus \underbrace{I(\xi) \oplus[-C_{\text{sys}}F_{\text{sys}}^{-1}J(\xi)]}_{S(\xi, s)}\\
&\Delta z \in C_w \Delta x_{w} + D_w u_{\text{sp}} \oplus \underbrace{K(\xi) \oplus [- N_{\text{sys}}F_{\text{sys}}^{-1}J(\xi)]}_{O(\xi, s)}
\end{aligned}
\end{align}
where the terms $S(\xi, s)$ and $O(\xi, s)$ bound the linearization errors.

\subsection{Numerical Evaluation of Linearization Errors}\label{sec_model_sub_WTG_error}
The linearization error terms $S(\xi, s)$ and $O(\xi, s)$ are complicated functions of several differential and algebraic variables besides the control inputs. It is extremely difficult to incorporate these terms in the optimization model analytically directly, and numerical methods are indispensable. Thus, we will first evaluate their numerical intervals, and incorporate them as the unknown-but-bounded uncertainty sets into the optimization. The subsection will describe essential steps to obtain the numerical intervals.

The first step is to enclose $\xi$ and $s$ in intervals. This step can be done by first computing the reachable sets of (\ref{eq_DAE}). With the time scale of interests, we can assume that the variations are caused only by the grid supportive control signals. Therefore, the over-approximation of the reachable sets can be performed if we bound the control input as $\mathcal{U}$, where $u_{\text{sp}}\in\mathcal{U}$. In turn, this control bound $\mathcal{U}$ will be enforced in the NOC. The over-approximation algorithms propagate the set-represented inputs and initial conditions under the system vector fields efficiently. The efficiency relies on the special representations of sets as boxes, ellipsoids, polytopes, support functions, and so on \cite{zhang_set_review}. Among all representations, the zonotope-based reachability computation has been applied to nonlinear differential-algebraic systems \cite{Althoff2014}. Then, the obtained reachable sets, which are typically high-order zonotopes and difficult for general computational operations, will be converted into intervals by certain interval hull over-approximation methods. A detailed description of the algorithms is out of the scope of this paper and can be found in \cite{Althoff2014}.
We employ the zonotope-based reachability analysis and perform the computation using the Matlab toolbox CORA \cite{Althoff2014}. Eq. (\ref{eq_xi_set}) indicates that $\xi$ and $s$ can be enclosed in the same interval. Let this interval be denoted as $\varTheta$, where $s,\xi\in\varTheta$.

The second step is to conduct the bound evaluation for functions $S(\xi, s)$ and $O(\xi, s)$ subjected to the interval inputs $s$ and $\xi$, where $s,\xi\in\varTheta$. This problem can be formulated as two nonlinear optimization problems. However, the results could be local optimum and fail the enclosure. Other approaches, such as Monte-Carlo sampling, are subjected to the similar flaw. On the other hand, the interval arithmetics, or interval analysis, aims to define a systematic rule of operations for intervals, interval vectors, and matrices such that the exact solutions are always included \cite{interval2006}. Therefore, we will employ the interval arithmetics to calculate the interval of $S(\xi, s)$ and $O(\xi, s)$, denoted as $\mathcal{S}$ and $\mathcal{O}$, respectively. Thereto, we first compute the intervals of every element in the Hessian matrices $H^{d}_{i}(\xi)$, $H^{a}_{i}(\xi)$ and $H^{o}_{i}(\xi)$ to form the interval Hessian matrices $\mathcal{H}^{d}_{i}$, $\mathcal{H}^{a}_{i}$ and $\mathcal{H}^{o}_{i}$, where the possible values of Hessian matrices are enclosed. Then, based on the results from the first step, we can express the term $\delta=s-s^{\text{eq}}$ as an interval vector, where $s\in\varTheta$. Performing the interval multiplication between the interval vector $\delta$ and the interval Hessian matrices yields the interval vectors of the Lagrangian remainders
\begin{align}
\label{eq_interval_Lagrang}
\begin{aligned}
\mathcal{I}_i=[\underline{I}_i,\overline{I}_i]=\frac{1}{2}\delta^{T}\mathcal{H}^{d}_i\delta\\
\mathcal{J}_i=[\underline{J}_i,\overline{J}_i]=\frac{1}{2}\delta^{T}\mathcal{H}^{a}_i\delta\\
\mathcal{K}_i=[\underline{K}_i,\overline{K}_i]=\frac{1}{2}\delta^{T}\mathcal{H}^{o}_i\delta\\
\end{aligned}
\end{align}
where $\underline{I}_i$ and $\overline{I}_i$ represent the upper and lower bound of $I_i(\xi)$ for $\xi\in\varTheta$, and the same notation is applied for $J_i(\xi)$ and $K_i(\xi)$ as well. At last, the interval representations of the linearization errors can be calculated as follows
\begin{align}
\label{eq_interval_error}
\begin{aligned}
\mathcal{S}=\mathcal{I}-C_{\text{sys}}F_{\text{sys}}^{-1}\mathcal{J}\\
\mathcal{O}=\mathcal{K} - N_{\text{sys}}F_{\text{sys}}^{-1}\mathcal{J}
\end{aligned}
\end{align}
Note that computations in (\ref{eq_interval_Lagrang}) and (\ref{eq_interval_error}) are interval operations defined in \cite{interval2006}. Such computations can be performed either by CORA \cite{Althoff2014}.


\section{Robust Numerical Optimal Control Synthesis}\label{sec_method}
\subsection{Overall Configuration}\label{sec_method_sub_overall}
\begin{figure}[h]
	\centering
	\includegraphics[scale=0.4]{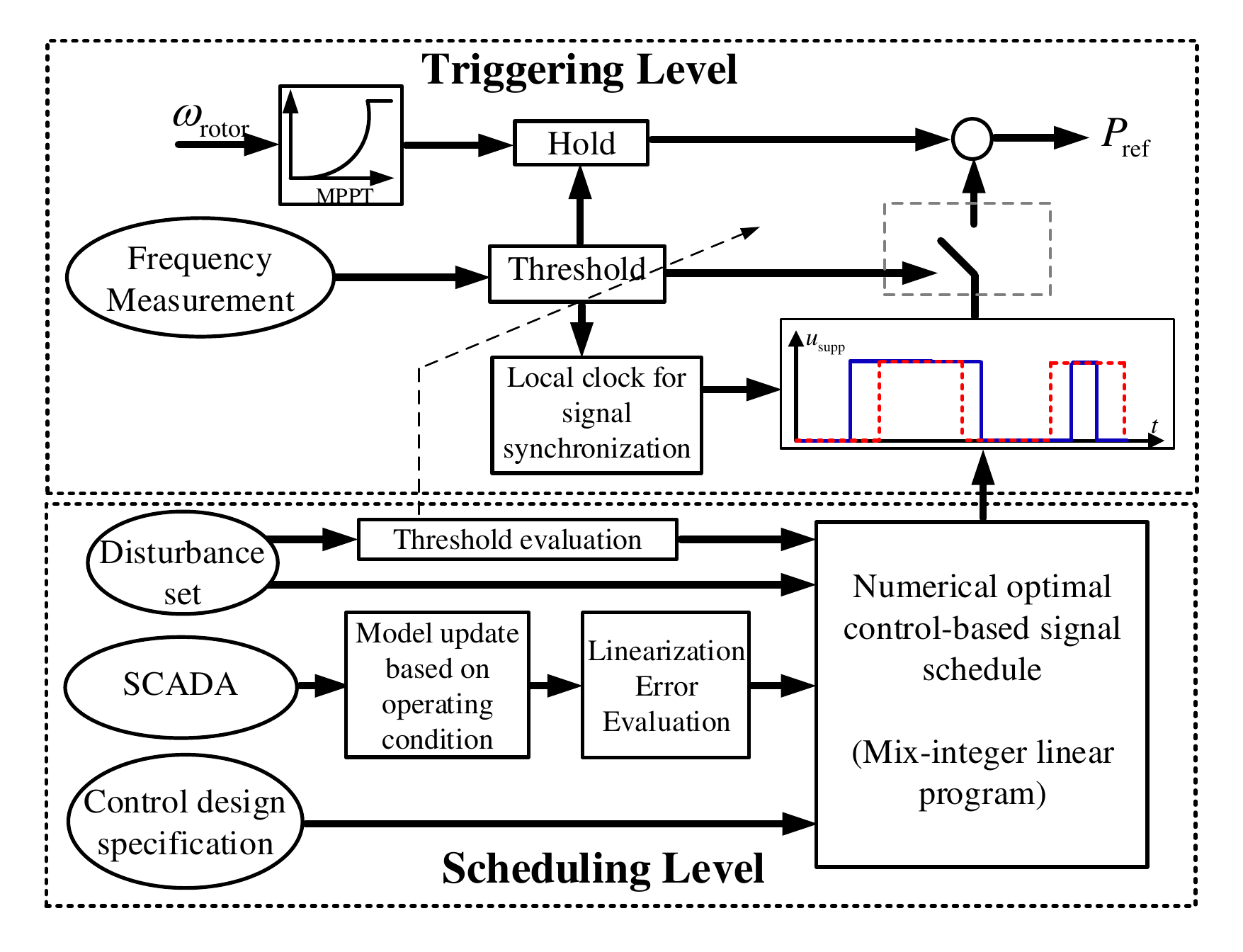}
	\vspace{-0.15in}
	\caption{Overall configuration of synthesizing performance guaranteed controller.}
	\label{fig_overall}
\end{figure}

The overall configuration of the proposed control is illustrated in Fig. \ref{fig_overall}. 
The objective is to preventively compute feasible supplementary control signals for WTGs to support the frequency within permissible ranges. 
The control paradigm is configured into two levels, that is, the upper centralized scheduling level and the lower decentralized triggering level.

In the scheduling level, a numerical optimal control (NOC) problem, described in Subsection \ref{sec_method_sub_formulation}, is formulated and solved for GS control inputs. To ensure that the control inputs are robust against varying operating conditions, the grid status is acquired to update the parameters of the NOC model, including model parameters and linearization errors. The linearization error analysis will be re-performed once the original operating points of WTGs change. 
With up-to-date information, the NOC problem formulated as a MILP can be promptly solved by off-the-shelf commercial solvers. 

The triggering level is configured locally at each WTG, where the up-to-date control signals are stored in the designated WTGs. Once the frequency deviation exceeds a certain threshold, stored control signals are synchronized with the real-time and applied to the supplementary loop of the WTGs. 

\subsection{Numerical Optimal Control for Scheduling Level}\label{sec_method_sub_formulation}
The AFR model can be obtained by integrating the WTG model in (\ref{eq_WTG_linear_2}) with the frequency response model of DG in (\ref{eq_SFR}) as illustrated in Fig. \ref{fig_Model_Full} (c). Without loss of generality, we consider a microgrid with one DG and multiple WTGs. The term $N_{w}$ is the number of WTGs, and $\mathcal{N}_{w}$ is the set of WTG indices. Let $S_{d}$ and $S_{w,i}$ be the base of DG and WTG $i$, respectively, and, $k_{d}=1/S_{d}$, $k_{dw,i}=S_{w,i}/S_{d}$. It is essential to note that the uncertainty is introduced only when the WTG is activated for the grid supportive mode. Therefore, we introduce the binary variable $b_{i}$ to indicate the activation status of WTG $i$, that is, $b_{i}$ equals to one when the grid-supportive mode is on and zero otherwise. With all the above definition, the AFR can be expressed as follows
\begin{align}
\label{eq_AFR}
\begin{aligned}
&2H_{d}\Delta\dot{\omega}_{d}\in \overline{f}(\Delta P_{m}+\sum_{i=1}^{N_{w}}k_{dw,i}\Delta P_{g,i})\oplus\overline{f}k_{d}\mathcal{P}_{\text{dis}}\\
&\tau_{d}\Delta\dot{P}_{m}\in -\Delta P_{m}+\Delta P_{v}\\
&\tau_{g}\Delta\dot{P}_{v}\in  -\Delta P_{v}  - \Delta\omega_{d}/(\overline{f}R_{D})\\
&\Delta\dot{x}_{w,i}\in  A_{w,i}\Delta x_{w,i} + B_{w,i}u_{s,i} \oplus b_{i}\mathcal{S}_i\quad i\in\mathcal{N}_{w}
\end{aligned}
\end{align}
where
\begin{align}
\label{eq_WTG_linear_power}
\begin{split}
&\Delta P_{g,i}\in C_{w,i}\Delta x_{w,i} + D_{w,i}u_{s,i} \oplus b_{i}\mathcal{O}_{i}\quad i\in\mathcal{N}_{w}
\end{split}
\end{align}
where $\mathcal{P}_{\text{dis}}$ is the predefined disturbance set. To obtain the overall state-space model, we first define the state and input vectors as follows
\begin{equation*}
\begin{aligned}
& x=[\Delta\omega_{d},\Delta P_{m},\Delta P_{v},\Delta x_{w,1},\cdots,\Delta x_{w,i},\cdots,\Delta x_{w,N_{w}}]^{T}\\
& u=[u_{\text{sp},1},\cdots,u_{\text{sp},i}\cdots,u_{\text{sp},N_{w}}]^{T}\\
& b=[b_{1},\cdots,b_{i}\cdots,b_{N_{w}}]^{T}\\
&\mathcal{S}=[\mathcal{S}^{T}_1,\cdots,\mathcal{S}^{T}_i,\cdots,\mathcal{S}^{T}_{N_{w}}]^{T}\\
&\mathcal{O}=[\mathcal{O}_1,\cdots,\mathcal{O}_i,\cdots,\mathcal{O}_{N_{w}}^{T}]^{T}\\
\end{aligned}
\end{equation*}
Then, substituting (\ref{eq_WTG_linear_power}) into (\ref{eq_AFR}) yields the following state-space model
\begin{align}
\label{eq_AFR_contin}
\dot{x}\in Ax + B_{1}u \oplus B_{2}k_{d}\mathcal{P}_{\text{dis}} \oplus B_{3}(b\circ\mathcal{S}) \oplus B_{4}(b\circ\mathcal{O})
\end{align}
where $\circ$ denotes the Hadamard product, which performs the element-wise multiplication of two matrices with the same dimensions. The analytical model in (\ref{eq_AFR_contin}) is discretized at a sample time of $t_{s}$ and expressed compactly as follows
\begin{align}
\label{eq_SS_discrete}
\begin{aligned}
x(k+1) \in &A_{d}x(k)+B_{d1}u(k) \oplus B_{d2}k_{d}\mathcal{P}_{\text{dis}} \\
&\oplus B_{d3}[b(k)\circ\mathcal{S}(k)] \oplus B_{d4}[b(k)\circ\mathcal{O}(k)]
\end{aligned}
\end{align}
Let $\mathcal{T}=[0,1,\cdots,T]$ be the discretized time series of the overall scheduling horizon, $k\in\mathcal{Z}=[1,2,\cdots,Z]$ be its indices, and $\varGamma(\bullet):\mathbb{R}\longmapsto\mathbb{Z}$ be the mapping from the time series to the indices. First, the frequency deviation, that is, the rotor speed of the DG, should not exceed a certain limit at any time, that is, 
\begin{align}
\label{eq_con_freq}
\Delta \omega_{d}(k)\leq \Delta f_{d,\text{lim}}\quad\forall k\in\mathcal{Z}
\end{align}
No absolute value is used since the reverse power flow at PCC is excluded based on real practice. Since the kinetic energy of WTGs will be transferred to active power to support the grid, the speed of WTGs will decrease from nominal values. This deviation is also desired to be limited for all WTGs
\begin{align}
|\Delta \omega_{r,i}(k)|\leq \Delta f_{w,\text{lim}}\quad\forall k\in\mathcal{Z},i\in\mathcal{N}_{w}
\end{align}
For the ease of implementation, we confine the control signal to be the multi-level step function. Therefore, the control inputs are subjected to the following constraints
\begin{align}
\begin{aligned}
u_{s,i}(k)=u_{i}(k)u_{\text{L}}\quad\forall k\in\mathcal{Z},i\in\mathcal{N}_{w}\\
0\leq u_{i}(k) \leq u_{\text{BD}}\quad\forall k\in\mathcal{Z},i\in\mathcal{N}_{w}
\end{aligned}
\end{align}
where $u_{i}$ is an integer variable indicating the power level of the grid-supportive mode of WTG $i$, $u_{\text{L}}$ is the fixed magnitude of one-level input, and $u_{\text{BD}}$ is the total number of power levels. Besides, we would like to limit the number of times that one WTG is activated for grid support. Thereto, we first build up the constraint between $b_{i}(k)$ and $u_{i}(k)$, that is, $b_{i}(k)$ equals to one if $u(k)_{i}> 0$ and zero if $u(k)_{i}=0$. This logic relation is expressed by the following constraints
\begin{align}
\begin{aligned}
0\leq -u_{i}(k)+Mb_{i}(k)\leq M-1\quad\forall k\in\mathcal{Z},i\in\mathcal{N}_{w}
\end{aligned}
\end{align}
where $M$ is a big positive number. Then, another binary variable $v_{i}(k)$ is defined to indicate the grid-supportive mode change from off to on of the WTG $i$ by enforcing $v_{i}(k)$ to be one if the grid-supportive mode of WTG $i$ is activated at time step $k$ and zero otherwise using the following constraint
\begin{align}
\begin{aligned}
v_{i}(k)\geq b_{i}(k)-b_{i}(k-1)\quad\forall k=2,\cdots,Z,i\in\mathcal{N}_{w}
\end{aligned}
\end{align}
Obviously, $v_{i}(k)$ will be enforced to equal to one if the grid-supportive mode of WTG $i$ is off at step $k-1$ and activated at step $k$. Otherwise, $v_{i}(k)$ could be either zero or one. Therefore, we impose the following constraints to limit the activation times of a WTG no more than two times during one event
\begin{align}
\begin{aligned}
\sum_{k=1}^{Z}v_{i}(k)\leq 2 \quad\forall i\in\mathcal{N}_{w}
\end{aligned}
\end{align}
The grid-supportive mode of all WTGs should be off at the beginning and end of the scheduling horizon
\begin{align}
\begin{aligned}
u_{i}(0)=0, u_{i}(Z)=0 \quad\forall i\in\mathcal{N}_{w}
\end{aligned}
\end{align}
Considering the threshold, the NOC should start from a designate initial condition instead of zero
\begin{align}
x(0)=X_{0}
\end{align}
The objective is to minimize the control efforts. The total control effort can be represented as the summation of all integer variables as
\begin{align}
C_{U}=\sum_{i=1}^{N_{w}}\sum_{k=1}^{T}u_{i}(k)
\end{align}
The scheduling problem can be summarized as follows

\begin{subequations}
\label{eq_NOC_summary}
\begin{align}
\min \quad & C_{U}=\sum_{i=1}^{N_{w}}\sum_{k=1}^{T}u_{i}(k)\\
\text{s.t.} \quad & x(0)=X_{0} \\
& x(k+1) \in A_{d}x(k)+B_{d1}u(k) \oplus B_{d2}k_{d}\mathcal{P}_{\text{dis}} \nonumber\\
&\quad\quad \oplus B_{d3}[b(k)\circ\mathcal{S}(k)] \oplus B_{d4}[b(k)\circ\mathcal{O}(k)] \\
& \Delta \omega_{d}(k)\leq \Delta f_{d,\text{lim}} && \hspace{-3cm} \forall k\in\mathcal{Z} \\
&|\Delta \omega_{r,i}(k)|\leq \Delta f_{w,\text{lim}} && \hspace{-3cm} \forall k\in\mathcal{Z},i\in\mathcal{N}_{w} \\
&u_{s,i}(k)=u_{i}(k)u_{\text{L}} && \hspace{-3cm} \forall k\in\mathcal{Z},i\in\mathcal{N}_{w} \\
&0\leq u_{i}(k) \leq u_{\text{BD}} && \hspace{-3cm} \forall k\in\mathcal{Z},i\in\mathcal{N}_{w} \\
&0\leq -u_{i}(k)+Mb_{i}(k)\leq M-1 \nonumber\\
& && \hspace{-3cm} \forall k\in\mathcal{Z},i\in\mathcal{N}_{w} \\
&\sum_{k=1}^{Z}v_{i}(k)\leq 2 && \hspace{-3cm} \forall i\in\mathcal{N}_{w} \\
&u_{i}(0)=0,u_{i}(Z)=0 && \hspace{-3cm} \forall i\in\mathcal{N}_{w}
\end{align}
\end{subequations}

\subsection{Robust Optimization Re-formulation}\label{sec_method_sub_robust}
Note that the problem (\ref{eq_NOC_summary}) is infinite-dimensional due to the uncertainty set. The formulation involving the uncertainty sets needs to be re-formulated so that the worse-case realization of the uncertainty can be revealed. Since the robust sets are interval and the problem is MILP, the standard re-formulation method in \cite{robust_opt,lofberg2003minimax,Bertsimas2006} can be employed. Particularly, dynamic systems with input uncertainty have been considered in \cite{lofberg2003minimax,Bertsimas2006}.

Let a realization of the set $\mathcal{P}_{\text{dis}}$, $\mathcal{S}(k)$ and $\mathcal{O}(k)$ at step $k$ be denoted as $\tilde{p}(k)$, $\tilde{s}(k)$ and $\tilde{o}(k)$, respectively. Then, the differential inclusion in (\ref{eq_SS_discrete}) can be expressed as
\begin{align}
\label{eq_linear_response_system}
\begin{aligned}
&x(k+1) =A_{d}x(k)+B_{d1}u(k)+B_{d2}k_{d}\tilde{p}(k) \\
&\quad\quad\quad\quad+ B_{d3}[b(k)\circ\tilde{s}(k)] + B_{d4}[b(k)\circ\tilde{o}(k)]
\end{aligned}
\end{align}
The constraints on evolution discrete-time dynamics can be expressed in a compact notation as
\begin{align}
\label{eq_mpc_compact}
\begin{aligned}
&\textbf{x}=\mathcal{A}x(0) + \mathcal{B}_{d1}\textbf{u} + \mathcal{B}_{d2}k_{d}\textbf{p} + \mathcal{B}_{d3}\textbf{s}+ \mathcal{B}_{d4}\textbf{o}
\end{aligned}
\end{align}
where
\begin{align*}
&\textbf{x}=\left[ x(1),x(2),\cdots,x(Z)\right]^T,\textbf{u}=\left[ u(0),u(1),\cdots,u(Z-1)\right]^T\\
&\textbf{p}=\left[ \tilde{p}(0), \tilde{p}(1), \cdots, \tilde{p}(Z-1)\right]^T\\
&\textbf{s}=\left[b(0)\circ\tilde{s}(0),b(1)\circ\tilde{s}(1),\cdots,b(Z-1)\circ\tilde{s}(Z-1)\right]^T\\
&\textbf{o}=\left[b(0)\circ\tilde{o}(0),b(1)\circ\tilde{o}(1),\cdots,b(Z-1)\circ\tilde{o}(Z-1)\right]^T
\end{align*}
\begin{align*}
&\mathcal{A}=\left[ \begin{array}{c} 
A\\
A^2\\
\vdots\\
A^N
\end{array} \right]
&\mathcal{B}_{di}=\left[ \begin{array}{cccc} 
B_{di} & 0 & \cdots & 0\\
AB_{di} & B_{di} & \cdots & 0\\
\vdots & \vdots & \ddots & \vdots\\
A^{N-1}B_{di} & A^{N-2}B_{di} & \cdots & B_{di}
\end{array} \right]
\end{align*}
where $i$ denotes the different subscripts of the input matrices. Substituting (\ref{eq_mpc_compact}) into constraint (\ref{eq_con_freq}) yields
\begin{align*}
-\mathcal{A}x(0) - \mathcal{B}_{d1}\textbf{u} - \mathcal{B}_{d2}k_{d}\textbf{p} - \mathcal{B}_{d3}\textbf{s}- \mathcal{B}_{d4}\textbf{o}\leq E\Delta f_{d,\text{lim}}
\end{align*}
where $E$ will ensure the constraint to be effective at appropriate rows. Grouping all uncertainty sets on the left-hand side yields 
\begin{align}
\label{eq_con_robust_sat_1}
-\mathcal{B}_{d2}k_{d}\textbf{p} - \mathcal{B}_{d3}\textbf{s}- \mathcal{B}_{d4}\textbf{o}\leq E\Delta f_{d,\text{lim}} + \mathcal{A}x(0) + \mathcal{B}_{d1}\textbf{u} 
\end{align}
The robust constraint sanctification of (\ref{eq_con_robust_sat_1}) can be formulated as  \cite{Bertsimas2006}
\begin{align}
\label{eq_con_robust_sat_2}
\begin{aligned}
&\max_{w}(-\mathcal{B}_{d2}k_{d}\textbf{p}- \mathcal{B}_{d3}\textbf{s}- \mathcal{B}_{d4}\textbf{o}   ) \leq E\Delta f_{d,\text{lim}} \\&\quad\quad+ \mathcal{A}x(0) + \mathcal{B}_{d1}\textbf{u} + \mathcal{B}_{d2}k_{d}P_{\text{pcc}}
\end{aligned}
\end{align}
Due to the special form of the uncertainty set, that is, interval, the worst-case realization can be revealed row-wise by determining the positivity of the entry of matrices $\mathcal{B}_{d2}$, $\mathcal{B}_{d3}$ and $\mathcal{B}_{d4}$. In the power outage scenario, the element-wise positivity of $\mathcal{B}_{d2}$ indicates that we only need to consider the largest outage in $\mathcal{P}_{\text{dis}}$, denoted as $\overline{P}_{\text{dis}}$. While inconsistent signs in $\mathcal{B}_{d3}$ and $\mathcal{B}_{d4}$ will realize both upper and lower bounds.

\section{Case Study}\label{sec_sim}

\subsection{Enclosing Linearization Error}\label{sec_sim_sub_error}
\begin{figure}[h]
	\centering
	\includegraphics[scale=0.23]{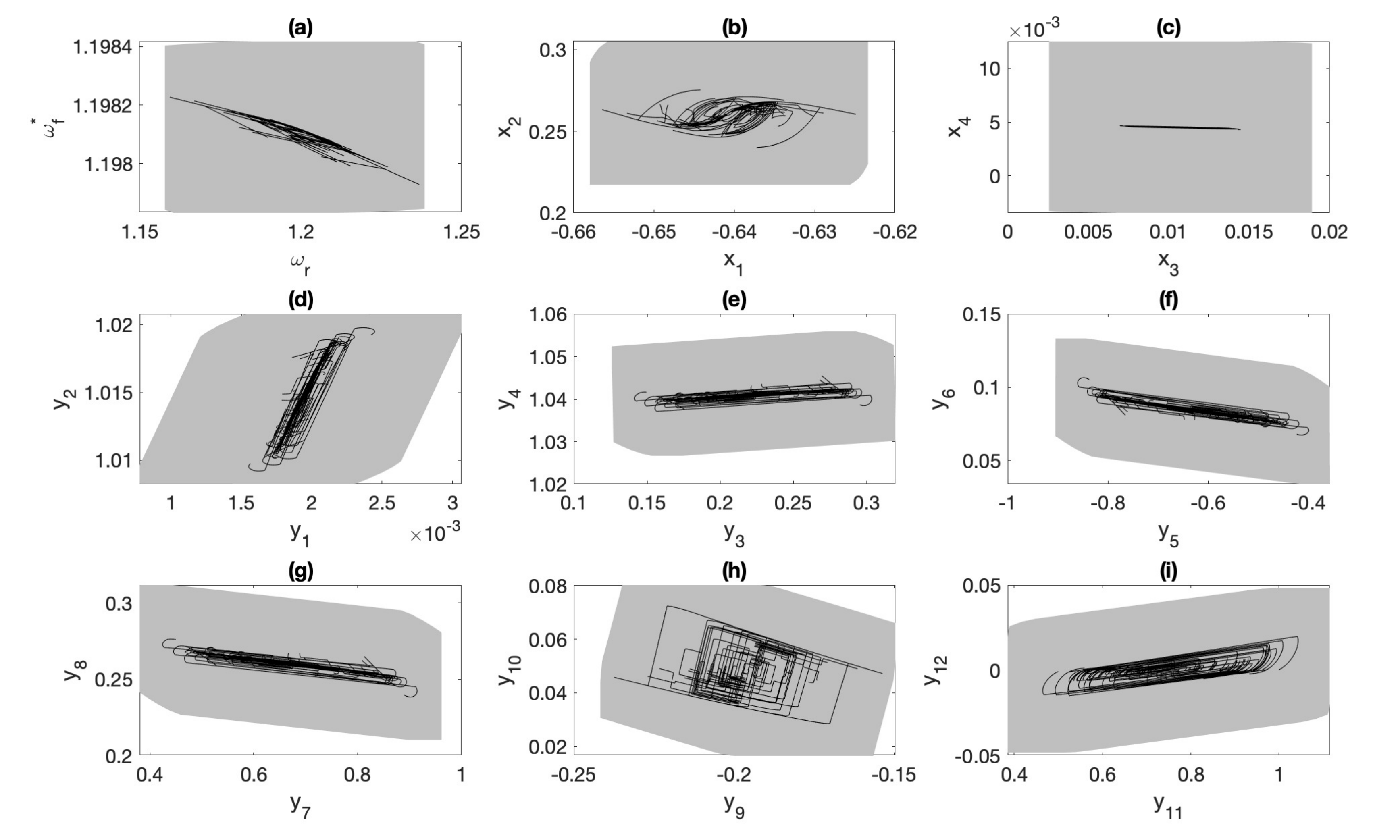}
	\vspace{-0.15in}
	\caption{Reachable sets of nonlinear WTG model under bounded input.}
	\label{fig_reach_nonlinear}
\end{figure}

In this subsection, we will demonstrate the results of the analysis presented in Section \ref{sec_model_sub_WTG_error}. As a prerequisite for the interval analysis, reachability analysis of the WTG in (\ref{eq_IM_ode5})-(\ref{eq_RSC_control_alg2}) is performed using CORA \cite{Althoff2014} under a bounded control input $\mathcal{U}=[0,0.1]$, which will be enforced in the NOC. The reachable sets are shown in Fig. \ref{fig_reach_nonlinear} in the format of zonotopes, which are further converted into intervals $\varTheta$. Then, the computations in (\ref{eq_interval_Lagrang}) and (\ref{eq_interval_error}) are performed to obtain the linearization error bounds in the format of intervals as follows
\begin{align*}
\mathcal{S}=\left[ \begin{array}{cc} 
&[-0.001192776663658, 0.001190937295999]\\
&[-0.000098120800673, 0.000098003228640]\\
&[0,0]\\
&[-0.056868144117605, 0.056840867582327]\\
&[-0.052150975958212, 0.052152050866097]\\
&[-0.104503795886190, 0.104504716861513]
\end{array} \right]
\end{align*}
\begin{align*}
\mathcal{O}=\left[ \begin{array}{cc} 
&[-0.024807136679364, 0.024828861362883]
\end{array} \right]
\end{align*}

\begin{figure}[h]
	\centering
	\includegraphics[scale=0.23]{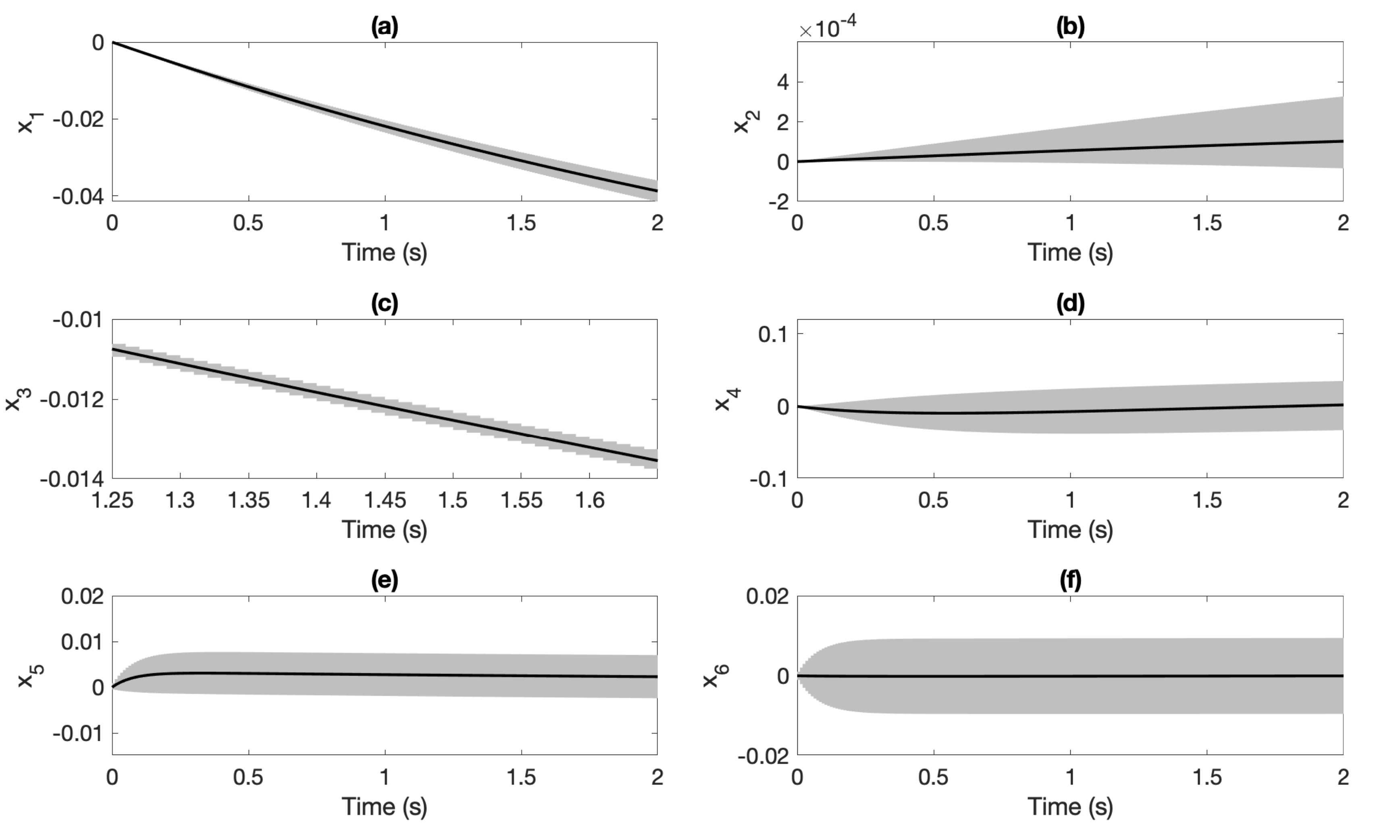}
	\vspace{-0.15in}
	\caption{Enclosing nonlinear trajectories by linear WTG with linearization error intervals. }
	\label{fig_reach_linear}
\end{figure}

To further verify the bounds, reachability analysis of the linearized WTG in (\ref{eq_WTG_linear_2}) under the inputs of a constant supplementary signal $u_{\text{sp}}$ and error sets $\mathcal{S}$ and $\mathcal{O}$. Simulated trajectory using the nonlinear model (\ref{eq_IM_ode5})-(\ref{eq_RSC_control_alg2}) under the same $u_{\text{sp}}$ is compared in Fig. \ref{fig_reach_linear}. With the obtained bounds, reachable sets of the linear system always enclose the nonlinear system's trajectory. Nate that the equilibrium points have subtracted the simulated nonlinear trajectories.


\subsection{Frequency Control During Islanding}
The optimization problem (\ref{eq_NOC_summary}) can be converted into a MILP after applying the technique in Section \ref{sec_method_sub_robust}. The problem is formulated in the Yalmip environment \cite{yalmip} and solved by efficient solvers Gurobi. The parameters in the MILP are given as follows
\begin{equation*}
\begin{split}
& t_{s}=0.1\text{ [s]},T=10\text{ [s]},u_{\text{L}}=0.02,u_{\text{BD}}=5\\
& \Delta f_{d,\text{lim}}=0.5\text{ [Hz]},\Delta f_{w,\text{lim}}=2\text{ [Hz]},\overline{P}_{\text{dis}}=0.7 [MW]
\end{split}
\end{equation*}
Based on the given parameters, it is required that the frequency deviation is limited within 0.5 Hz. 

\begin{figure}[h]
	\centering
	\includegraphics[scale=0.25]{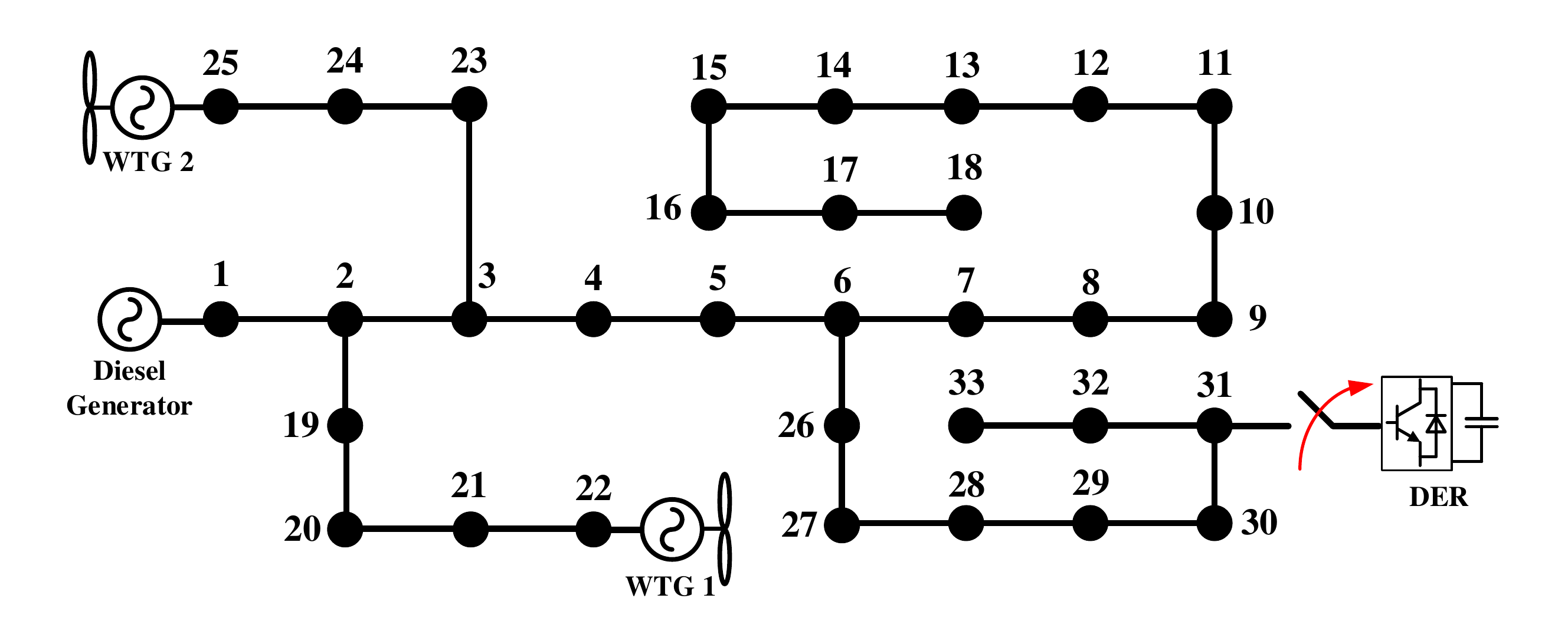}
	\vspace{-0.1in}
	\caption{Modified 33-node microgrid with diesel and wind turbine generators.}
	\label{fig_Model_Network}
\end{figure}

\begin{figure}[h]
	\centering
	\includegraphics[scale=0.23]{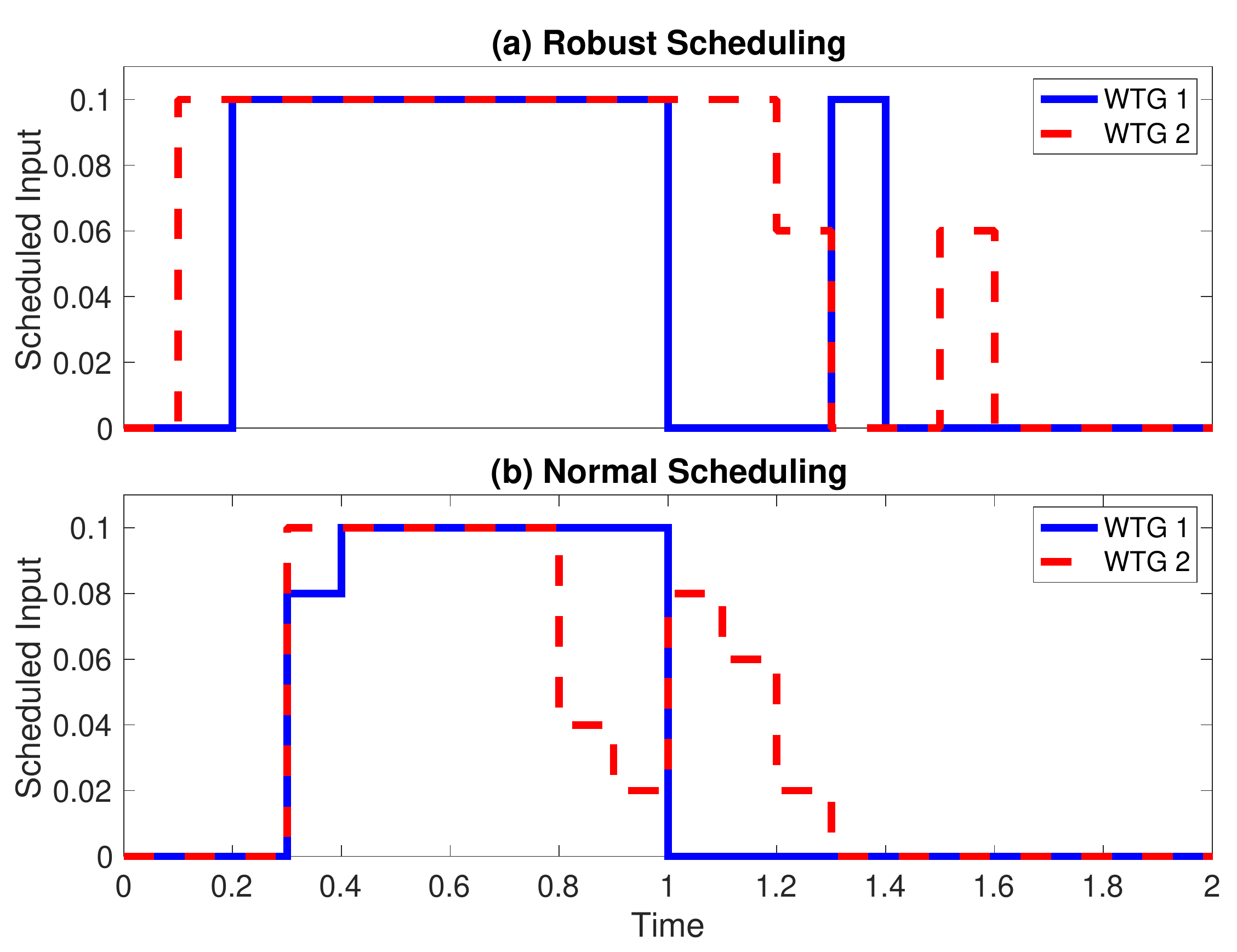}
	\vspace{-0.1in}
	\caption{Modified 33-node microgrid with diesel and wind turbine generators.}
	\label{fig_control_signal}
\end{figure}

The microgrid model for verification is a modified 33-node three-phase built in Simulink environment illustrated in Fig. \ref{fig_Model_Network}. All components shown in Fig. \ref{fig_Model_Full} have been implemented. Once the control signals are computed, they are set in the Simulink environment.  Cascade-connected switches are employed to realize the control signal, the input magnitude, and switching time of which will be adjusted accordingly.
We consider the worst-case disturbance to be the tripping of a distributed energy resource (DER) at its maximum power rating, that is, 0.7 MW, illustrated in Fig. \ref{fig_Model_Network}. The initial condition is set to be $X_{0}=[-0.11,0,0,0,\cdots,0]^{T}$ under current worse-case disturbance. After completing the up-to-date information process, the problem (\ref{eq_NOC_summary}) is solved less than 10 seconds. The result is plotted in Fig. \ref{fig_control_signal} (a). The NOC problem without considering uncertainty is also solved and plotted in \ref{fig_control_signal} (b). The total control effort $C_{U}$ with the presence of the uncertainty is 42, while without considering uncertainty is 33. In addition, since the uncertainty is only presented when the WTG is activated for grid support, the control signals tend to attain their maximum values. In the latter case, though the grid-supportive mode activation duration is approximately the same, the signals do not reach their output limits in most of the activation duration.

\begin{figure}[htbp!]
	\centering
	\includegraphics[scale=0.3]{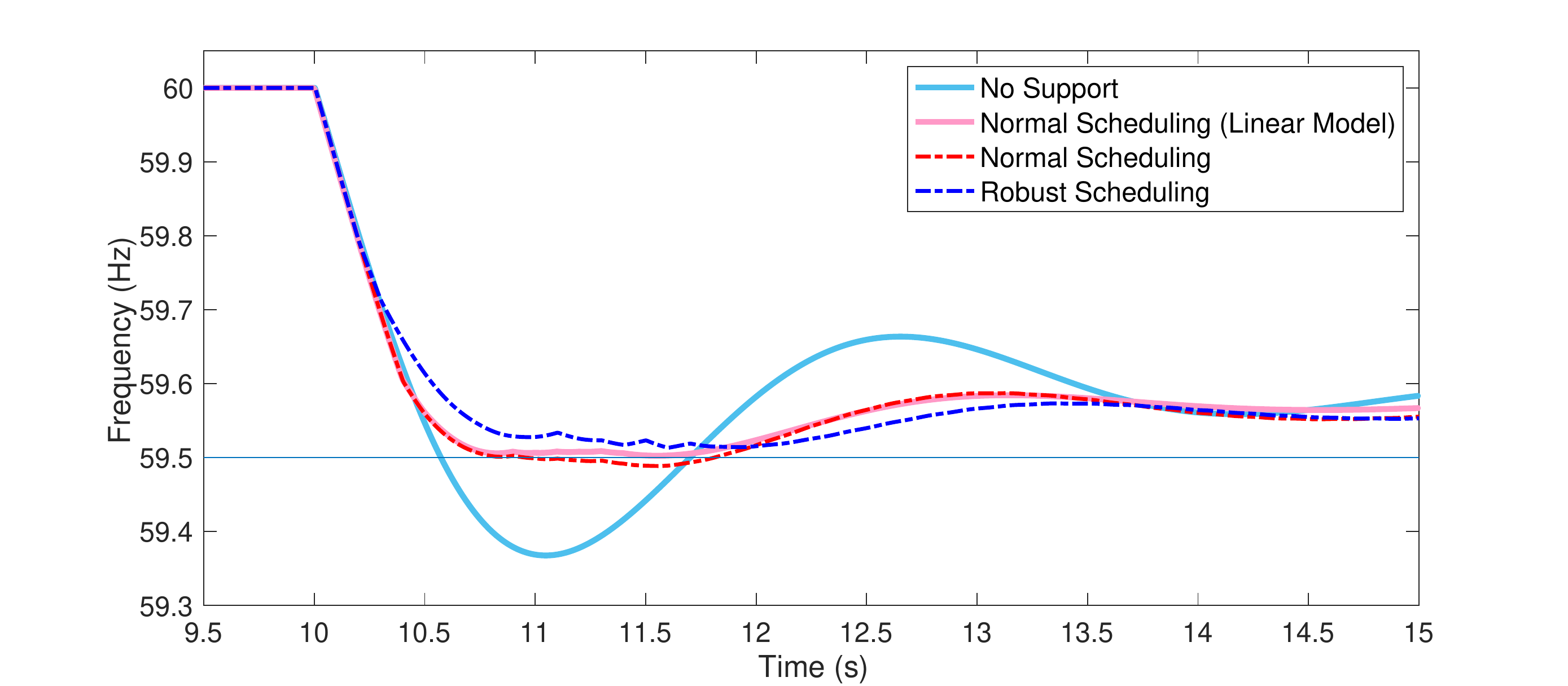}
	\vspace{-0.1in}
	\caption{Frequencies of DG under different cases.}
	\label{fig_Freq}
\end{figure}

\begin{figure}[htbp!]
	\centering
	\includegraphics[scale=0.3]{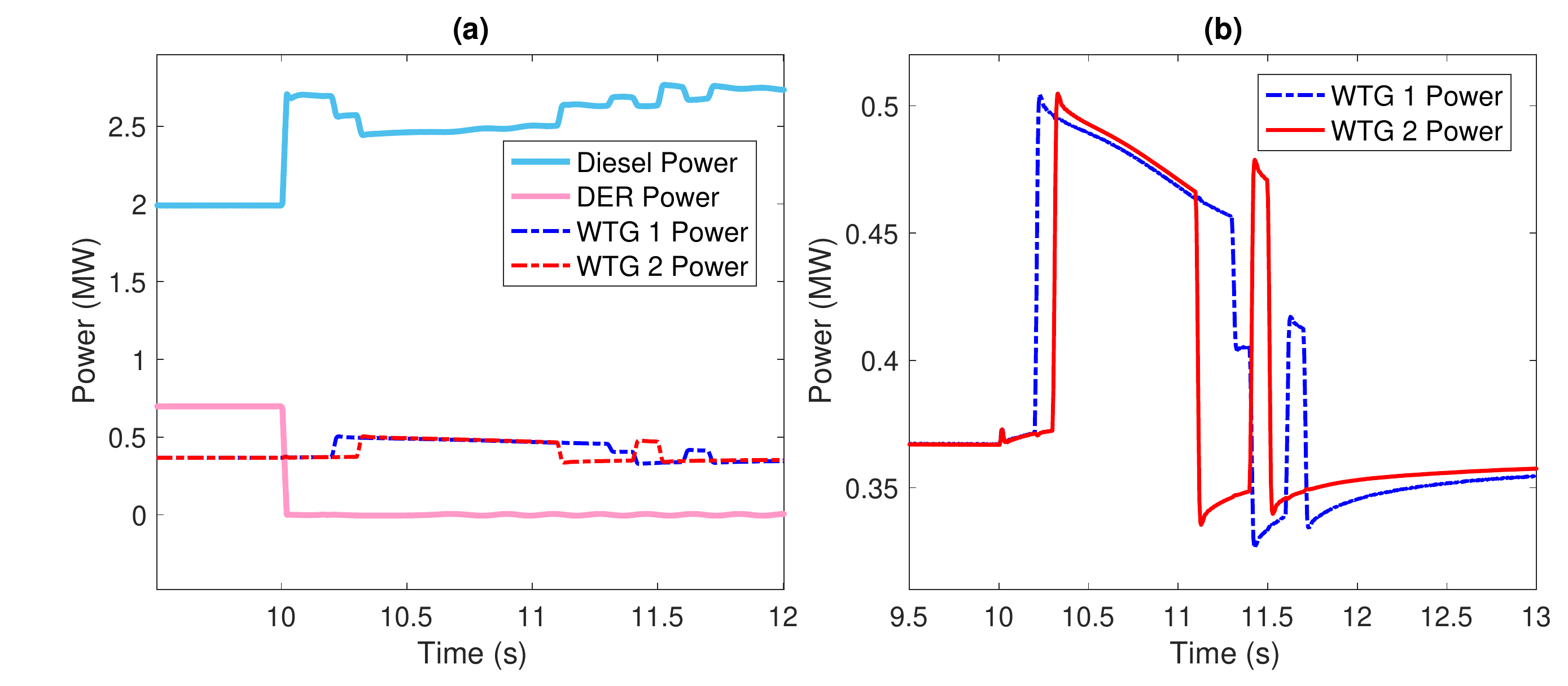}
	\vspace{-0.1in}
	\caption{(a) Active power in the network. (b) Enlarged view of WTG power.}
	\label{fig_Power}
\end{figure}

\begin{figure}[htbp!]
	\centering
	\includegraphics[scale=0.3]{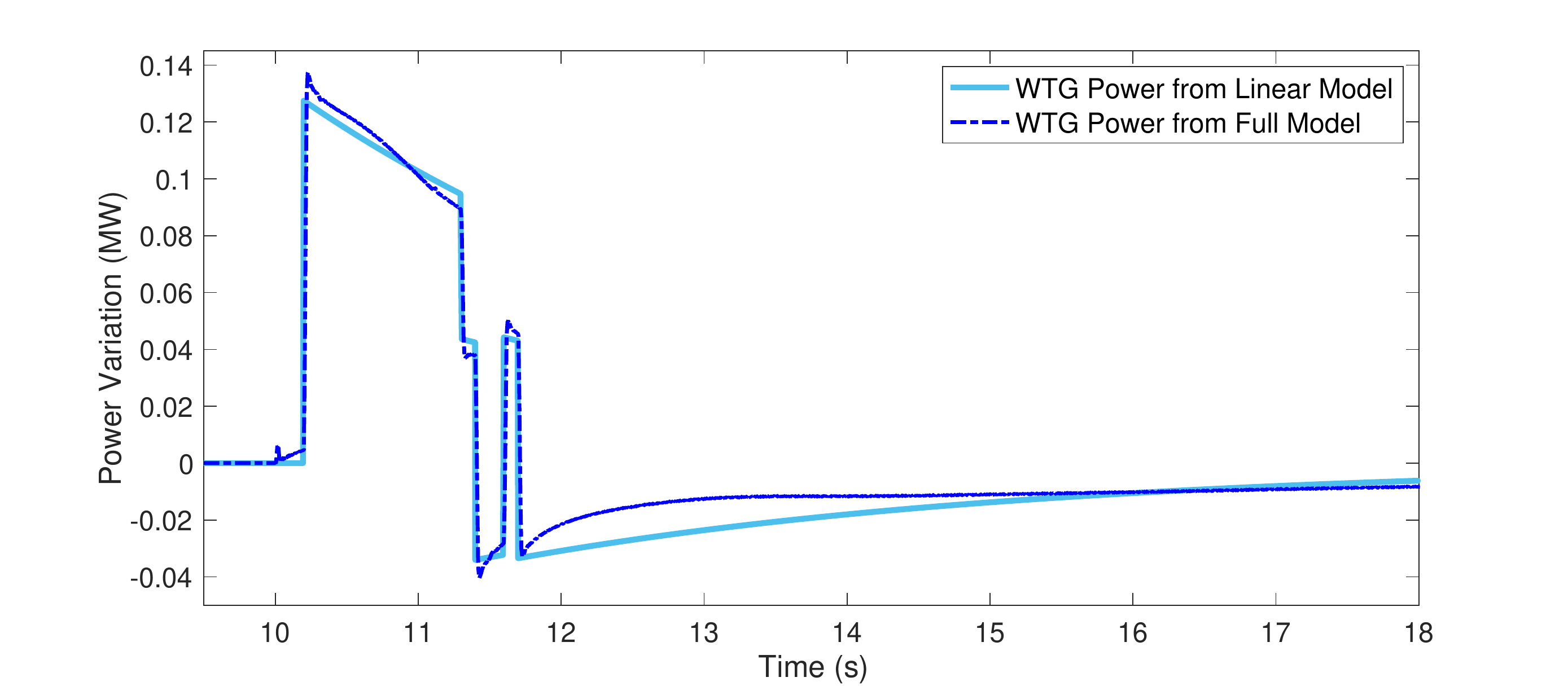}
	\vspace{-0.1in}
	\caption{Active power variations of WTGs from the linearized and full nonlinear model.}
	\label{fig_WTG_power}
\end{figure}

The computed signals are then equipped in the full-order three-phase nonlinear Simulink model for verification.  We disconnect the DER to simulate the worst-case disturbance. Once the frequency deviations cross the threshold $X_{0}$, the GS modes of WTGs are activated. The frequency responses of DG under no support, normally scheduled support, and robustly scheduled support are shown in Fig. \ref{fig_Freq}. The active power of the network is shown in Fig. \ref{fig_Power}. The previous two cases do not lead to safe responses, while the frequency in the last case stays within the permissible limits. In addition, the normally scheduled control is also applied to the linear system in Eq. (\ref{eq_linear_response_system}) without considering the uncertainty, and the response is shown in Fig. \ref{fig_Freq} for comparison. As we can see, this response is safe but leaving no extra margins since the control effort is to be minimized. Therefore, if the linearization error of WTGs is not taken into account, the frequency response losses its safety. The active power variations of WTG 1 from linear and full-order nonlinear models are shown in Fig. \ref{fig_WTG_power}.

\section{Conclusions} \label{sec_conclusion}
In this paper, a NOC-based control synthesis methodology is proposed for microgrid frequency control that can take the trajectory constraints into account. The key feature of the proposed paradigm is near real-time centralized scheduling for real-time decentralized executing. The controller schedules ahead a series of control signals to synthesize the grid-supportive mode of WTGs by solving the NOC problem, where the frequency response predicted by the AFR model satisfies the defined specifications under the predefined disturbance set. Then, the computed signals are transmitted to individual WTGs for local activation. Linearization-induced uncertainty of WTGs is derived and computed using interval analysis. The proposed control is verified on the full nonlinear model in Simulink. The simulation results indicate the scheduling control can successfully retain the frequency within permissible ranges under the islanding event. Particularly, case studies show that considering uncertainty can create extra safety margins for accommodating modeling errors and reserve safe responses. On the other hand, an unsafe trajectory occurs if the uncertainty is omitted.

\bibliographystyle{IEEEtran}
\bibliography{IEEEabrv_zyc,library,Ref_Robust}

\end{document}